\documentclass[letterpaper, showpacs,floatfix, aps,prl,amsmath,twocolumn,superscriptaddress]{revtex4}

\pdfoutput=1

\usepackage{epsfig}
\usepackage{verbatim}
\usepackage{epsf}
\usepackage{array}

\newcommand{\be}{\begin{equation}}
\newcommand{\ee}{\end{equation}}
\newcommand{\bea}{\begin{eqnarray}}
\newcommand{\eea}{\end{eqnarray}}
\newcommand{\w}{\omega}

\newcommand{\e}{\epsilon}

\newcommand{\St}{\widehat{\mathrm{St}}}
\renewcommand{\Re}{\mathrm{Re}}

\newcommand{\req}[1]{Eq.~(\ref{#1})}
\newcommand{\reqs}[1]{Eqs.~(\ref{#1})}
\newcommand{\rref}[1]{(\ref{#1})}

\newcommand{\ocite}[1]{Ref.~\cite{#1}}

\newcommand{\gl}{\mathrm{N}_{\rm l}}
\newcommand{\gr}{\mathrm{N}_{\rm r}}
\newcommand{\gch}{\mathrm{N}_{\rm t}}

\newcommand{\mls}{\delta_1}

\begin{document}

\title{Effect of Coulomb interaction on current noise in open quantum dots}

\author{G. Catelani}
\affiliation{Laboratory of Atomic and Solid State Physics, Cornell University,
Ithaca, NY 14853}
\author{M.G. Vavilov}
\affiliation{Department of Physics, University of Wisconsin, Madison,
WI 53706}

\begin{abstract}
We analyze the effect of Coulomb interaction on
the noise of electric current through an open quantum
dot. We demonstrate that the ensemble average value of the noise
power acquires an interaction correction even for a dot coupled to
the leads by reflectionless point contacts, when the ensemble
average conductance is known to have no interaction corrections.
To leading order, the correction to the noise originates from the formation of a
nonequilibrium state of the Coulomb field describing the
interaction between electrons. We find the dependence of the
current noise power on the electron temperature, the applied voltage
bias, and the strength of the Coulomb interaction.
\end{abstract}
\date{\today}
\maketitle

The Coulomb interaction in quantum dots is usually associated with the
phenomenon of the Coulomb blockade~\cite{ABG}, which
occurs in a quantum dot coupled to the leads
through tunnel barriers. In the Coulomb blockade
regime, electric current strongly depends on
gate voltages due to the electrostatic energy cost to change the
electron number in the dot. As contacts
between the leads and the dot become more transparent, the Coulomb
blockade becomes less pronounced. In the case of fully transmitting
contacts (reflectionless contacts), the Coulomb interaction has no
effect on the conductance of a quantum dot~\cite{PRL82BA,PRB69GZ,PRL94BLF},
averaged over the unitary ensemble (i.e., in the
presence of magnetic field), but it still manifests itself in the
correlation functions of the conductance~\cite{PRB57AG}.

Additional information about electron correlations
can be gained by measuring other transport properties such as the shot
noise~\cite{ButtikerBlanter-PReps}.
Indeed, the study of the current noise in quantum dots is currently the
subject of numerous
experiments~\cite{PRL80iannaccone,PRL86oberholzer,PRB73CW,PRL97CW,PRL96barthold,PRL96kouwenhoven,0703419zhang},
aimed at understanding the interplay between Fermi statistics and
electron repulsion. This study may present opportunities to use this
interplay in quantum information technologies. In most cases
the experiments were performed in the regime of Coulomb blockade,
when coupling between leads and the dot is weak. The question arises
as to what happens to the power of the current noise as the contacts
become transparent. In this
paper, we investigate the effect of the Coulomb interaction on the
noise power of electric current, averaged over a unitary ensemble,
for quantum dots with reflectionless contacts.
In particular, we are interested in understanding how relaxation in
the dot affects the noise power.

A phenomenological description of the effects of interaction between
electrons on the current noise through a quantum dot was developed
within a model of ``fictitious voltage probes.''~\cite{PRB46BB} The
properties of such fictitious probes are assumed to
represent the proper effect of interactions on electrons, and three
main scenarios~\cite{9611140dJB,ButtikerBlanter-PReps} were
considered: (i) quasielastic scattering (phase relaxation), (ii)
inelastic scattering (energy relaxation), and (iii) heating. Although
this description is sufficient in many cases to characterize the
interaction effects on current noise, it does not specify the
relation between microscopic parameters of the system and the
magnitude of the current noise.
An alternative description of the interaction effects on current
noise was developed in Refs.~\cite{PRB67BN,PRL94BN,PRB72TN}, based on the
``effective action'' analysis~\cite{PhysRep198SZ} of the cumulant
generating function.
The effect of inelastic scattering was considered, for quantum contacts,
in Ref.~\cite{PRB72TN};
however, the configuration of the environment
was taken to be the equilibrium state of the field, decoupled from the
nonequilibrium electron system. Also, according to Ref.~\cite{PRL94BN}
the elastic interaction correction to the generating function vanishes for a quantum dot
with reflectionless contacts for any state of the dot; the effect of the electron-electron
interaction in the inelastic channel was not analyzed.

Here we show that, due to inelastic processes, in the stationary nonequilibrium
state of the dot's collective excitations a finite correction to the average current
noise arises, which survives even in the case of reflectionless channels, when the elastic
contribution vanishes.
For example, for a unitary ensemble of quantum dots coupled to the left (right)
electron reservoirs by point contacts containing $\gl$ ($\gr$)
reflectionless channels ($\gch=\gl+\gr \gg 1$), we obtain for the power of the
shot noise at zero temperature $T=0$ and intermediate bias
$1/\tau \ll |eV | \ll E_C \gch$ (here and below $\hbar=1$ and $k_{\rm B}=1$)
\be
S=2 e {\cal F} |I| \left[1+\frac{1-2{\cal F}}{\gch}\left(\ln|eV\tau|-1\right)\right],
\label{eq1}
\ee
where $I={\cal G}V$ is the current in response to voltage $V$,
${\cal G}=(e^2/2\pi\hbar)\gl\gr/\gch$ and ${\cal F}=\gl\gr/\gch^2$
are the ensemble average conductance and Fano factor, respectively.
The second term in \req{eq1} represents the $1/\gch$ correction due
to the Coulomb interaction,
$E_C=e^2/2C$ is the charging energy, $\tau=2\pi/(\gch\mls)$ is the electron dwell time in the dot,
and $\mls$ is the mean level spacing in the dot. As we employ the random matrix theory
description of the dots' ensemble, we assume that the Thouless energy $E_{T}=1/\tau_{\rm f}$,
where $\tau_{\rm f}$ is the electron flight time across the dot, is the largest energy scale,
so that, for example, $E_{T} \gg 1/\tau$.

We consider a quantum dot, coupled by reflectionless point contacts
to the electron reservoirs. Electron energy distribution is given by
$\tilde n_{\rm l}=n_{\rm F}(\e-eV)$ in the left reservoir and by
$\tilde n_{\rm r}=n_{\rm F}(\e)$ in the right reservoir, where
$n_{\rm F}(\e)=1/[\exp(\e/T)+1]$ is the Fermi distribution
function.  Taking into account the interaction between electrons in a quantum dot,
we obtain~\cite{cv-unp} the following expression for the ensemble averaged
noise power of the current through a quantum dot with reflectionless point contacts
in terms of electron distribution functions in
the leads $\tilde n_{\rm l,r}(\e)$ and in the dot $n(\e)$:
\be
\label{Sgen}
\begin{split}
S&  = 2{\cal G} \int\!d\e \ \bigg[  n(\e)(1-n(\e))
\\
& + \frac{\gr}{\gch} \tilde{n}_{\rm l}(\e)(1-\tilde{n}_{\rm l}(\e))
+ \frac{\gl}{\gch} \tilde{n}_{\rm r}(\e)(1-\tilde{n}_{\rm r}(\e))
\bigg],
\end{split}
\ee
We note that \req{Sgen} has the
structure identical to that for the noise power in the absence of
interaction. The interaction effects are completely hidden in the
configuration of the distribution function $n(\e)$ in the dot, and
the further evaluation of the interaction corrections to the noise reduces
to calculation of $n(\e)$.

The distribution function $n(\e)$ of electrons in the dot is found by
solving a kinetic equation. The derivation of the kinetic
equation for electrons in quantum dots can be found in \ocite{aca}.
Here we just present the explicit form of the kinetic equation for
the quantum dot with reflectionless contacts in a steady
state:
\be
\label{kineq}
\begin{split} \frac{n(\e)}{\tau}  = &
\frac{\mls}{2\pi}\Big[\gl \tilde{n}_{\rm l}(\e) + \gr \tilde{n}_{\rm r} (\e) \Big]
\\ & + \St_\mathrm{(in)}\{n,\tilde{n}_n,N^\alpha\} +
\St_\mathrm{(vb)}\{n,\tilde{n}_n\}.
\end{split}
\ee
The collision integral in the right hand side of \req{kineq}
is written as the sum of three terms. The first term  describes
the relaxation of $n(\e)$ due to the electron exchange between the leads and
the dot. The other two terms of the collision integral,
$\St_\mathrm{(in)}\{n,\tilde{n}_n,N^\alpha\}$ and
$\St_\mathrm{(vb)}\{n,\tilde{n}_n\}$, represent the effect
of electron-electron interaction in the Coulomb channel. We identify
$\St_\mathrm{(in)}\{n,\tilde{n}_n,N^\alpha\}$ as the inelastic
collision integral, corresponding to the processes in which an electron
changes its energy due to emission or absorption of a bosonic
collective excitation, while $\St_\mathrm{(vb)}\{n,\tilde{n}_n\}$ takes into
account electron-electron scattering via the exchange of virtual bosons.

The inelastic collision integral
\be
\label{inci}
\begin{split}
\St_\mathrm{(in)}(\e)& =  \frac{\mls^2}{4\pi^2} \sum_{n={\rm l,r}}
\mathrm{N}_n \int\!\frac{d\w}{\w} \\
 \times&
\Re \left\{
(1+F^\rho)
{\cal L}^\rho_\w \tilde{\Upsilon}^\rho_n(\e,\w)-
{\cal L}^g_\w \tilde{\Upsilon}^g_n(\e,\w)\right\}
\end{split}
\ee
can be written as a difference between a
charge boson part ($\rho$) and a ghost part ($g$).
The details about the need and meaning of this separation can
be found in Ref.~\cite{ca}. The propagators ${\cal L}^\alpha_\omega$ and coupling
constants $F^{\rho,g}$ are given by
\be
{\cal L}^\alpha_\omega = \frac{1}{-i\w + \left(1+F^\alpha\right)/\tau}\, ;
\ \ \  F^\rho = \frac{4E_c}{\delta_1}\, ; \quad F^g=0.
\ee
The functions $\tilde{\Upsilon}^\alpha_n(\e,\w)$ describe the differences
between absorption and emission rates of bosonic excitations
in lead $n=\{{\rm l,r}\}$:
\be
\label{upsdef}
\begin{split}
\tilde{\Upsilon}^\alpha_n(\e,\w) =
& \big( N^\alpha_\w + 1\big) \tilde{n}_n(\e) \big(1-\tilde{n}_n(\e-\w)\big)
\\ & - N^\alpha_\w\big(1-\tilde{n}_n(\e)\big)\tilde{n}_n(\e-\w).
\end{split}
\ee
The charge and ghost boson occupation numbers $N^{\rho, g}_\w$
satisfy the steady-state kinetic equations \cite{aca}
\be
0=\frac{1+F^\alpha}{2\w} \sum_{n={\rm l,r}} \mathrm{N}_n \int\!d\e
\left[ \tilde{\Upsilon}_n^\alpha (\e,\w) + \Upsilon^\alpha(\e,\w) \right]
\label{kinN}
\ee
with $\Upsilon^\alpha(\e,\w)$ obtained by replacing $\tilde{n}_n \to n$ in \req{upsdef}.
These equations give $N^{\rho, g}_\w=N_\w$ as functionals of $n(\e)$ and
$\tilde{n}_n(\e)$:
\be
\label{Nsol}
\begin{split}
&N_\w  =\frac{1}{2\w}\int\!d\e \sum_{n={\rm l,r}} \frac{{\mathrm N}_n}{\gch}
\\
& \times
\Big[ n(\e)\big(1-n(\e-\w) \big)
+ \tilde{n}_n (\e) \big(1-\tilde{n}_n(\e-\w)\big) \Big].
\end{split}
\ee
Note that the identity of the charge and ghost boson occupation numbers
also results in
$\tilde{\Upsilon}^{\rho, g}_n(\e,\w)=\tilde{\Upsilon}_n(\e,\w)$ and
$\Upsilon^{\rho, g}(\e,\w)=\Upsilon(\e,\w)$.

The second collision integral in \req{kineq} describes electrons interacting
by exchanging virtual bosons:
\be
\St_\mathrm{(vb)} = -\frac{\delta_1^2\gch}{2\pi^2}\int\!
\frac{d\w}{\w} \, \frac{\Upsilon(\e,\w)}{2+F^\rho}
\Re\left[(1+F^\rho){\cal L}^\rho_\w - {\cal L}^g_\w \right].
\label{civb}
\ee
We emphasize that this form of $\St_\mathrm{(vb)}$ given in terms of
$\Upsilon(\e,\w)$, which contains the boson occupation
number $N_{\w}$, is valid only in the steady state of the boson
fields. In a general time-dependent state, $\St_\mathrm{(in)}$, \req{inci},
contains the boson occupation numbers $N^{\rho, g}_\w$, which can be found as
solutions to the corresponding non-stationary counterpart of \req{kinN}, while
$\St_\mathrm{(vb)}$ is still given in terms of the function $\Upsilon(\e,\w)$
in which $N_\w$ is \emph{defined} by \req{Nsol} as a combination of electron
distribution functions $n(\e)$ and $\tilde n_{\rm l,r}(\e)$.

We look for an iterative solution to the kinetic equation \req{kineq} in
the form
\be
\label{n0} n(\e)=n_0(\e)+\delta n(\e)\,; \ \ \
n_0(\e)=\frac{\gl \tilde{n}_{\rm l}(\e)+\gr\tilde{n}_{\rm r}(\e)}{\gch},
\ee
where $\delta n(\e)$ represents the lowest order correction  to the distribution
function due to the Coulomb interaction. Here we assume that the dominant
scattering mechanism is the escape from the dot; therefore $\delta n$ contains the
additional small factor $1/\gch$. As a result, we can write the
expression for the power of low-frequency current noise in the form
\be
\begin{split}
& S=S_0+\delta S; \ \ \ S_0=4{\cal G}T+
{\cal G}  {\cal F} Y(eV,0);\\
& \delta S = 2{\cal G}
\int\!d\e \, \delta n(\e) \left(1-2n_0(\e)\right) \ .
\end{split}
\label{noise}
\ee
Here $S_0$ is the known result for the ensemble average value of
current noise power in the absence of interaction.
We introduced the function
\be
Y(U,\omega)=
\sum\limits_\pm
(U\pm \omega)\coth\frac{U\pm\omega}{2T}-
2\omega\coth\frac{\omega}{2T}\, ,
\label{Y}
\ee
which determines the power of the shot noise at frequency
$\omega$ through a point contact~\cite{PRL80SchProber}.
For the low-frequency noise power $S_0$ only the limit $Y(U,\omega\to 0)$ is
needed, but $Y(U,\omega)$ for arbitrary $\omega$
will also appear below.

\begin{figure}
\includegraphics[width=0.45\textwidth]{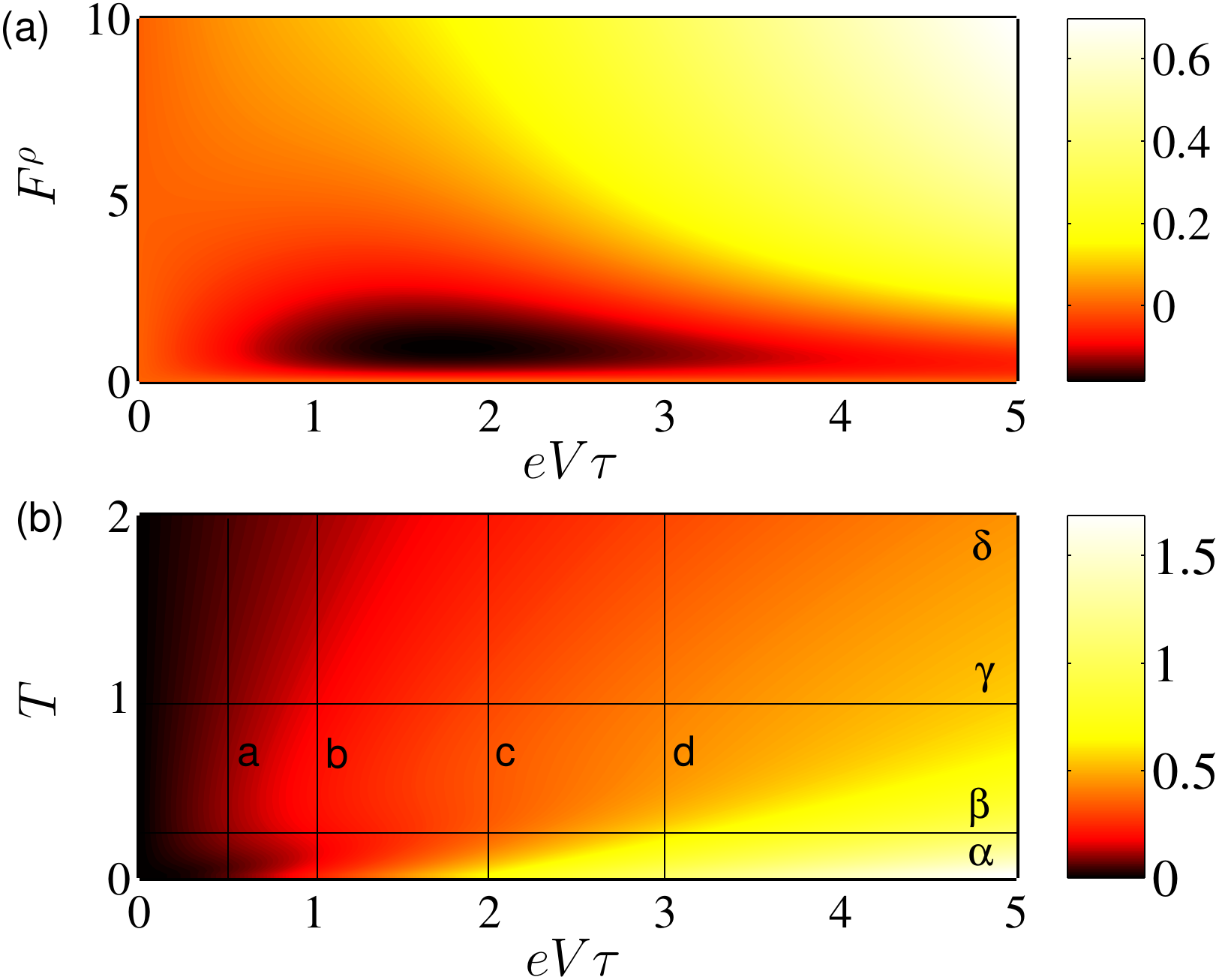}
\caption{(Color online). The correction to the noise power $\chi$ (a) as a function of bias and charging energy
for $T=1/\tau$ and (b) as a function of bias and temperature for $F^\rho =100$.}
\label{fig:1}
\end{figure}

To calculate the lowest order contribution to $\delta n(\e)$ due
to the Coulomb interaction, we replace  $n(\e)$ by $n_0(\e)$
in \reqs{Nsol} and \rref{civb}. As a result, we find the boson occupation
numbers
\be
N_\w = N^{\rm P}_\w + \frac{{\cal F} Y( eV,\w)}{4\w}\,  , \ \ \
N^{\rm P}_\w = \frac{1}{e^{\w/T}-1} \, .
\label{Nalpha}
\ee
Here, the first term is the equilibrium value of $N_\w$, represented
by the Planck function $N^{\rm P}_\w$, while the second term describes the
nonequilibrium contribution due to finite bias $V$.
The connection between $N_\w - N^{\rm P}_\w$ and the power of shot noise at
frequency $\w$ is not occasional, as the noise measurement at frequency $\omega$
has the physical interpretation~\cite{JMatPhys37LLL} of the
number of electromagnetic quanta detected by the measurement
circuit.

Substituting $n_0(\e)$, \req{n0}, and $N_\omega$, \req{Nalpha}, into
\reqs{inci} and \rref{civb}, we find a correction to the
electron distribution function $\delta n(\e)$
as the first iteration solution to the kinetic equation, \req{kineq}:
\be\label{itsol}\begin{split}
& \delta n(\e) =  {\cal F}\int\!\frac{d\w}{\w} \frac{\delta_1}{2\pi} \Re \left[(1+F^\rho){\cal L}^\rho_\w -
{\cal L}^g_\w \right] \\
& \qquad \bigg\{ \frac{F^\rho}{2+F^\rho} \frac{Y(eV,\w)}{4\w} \left[n_0(\e)-n_0(\e-\w) \right] \\
& -\frac{2}{2+F^\rho} \Big[\left(N^{\rm P}_\w - N^{\rm P}_{\w-eV}\right)
\left(\tilde{n}_{\rm l}(\e) - \tilde{n}_{\rm r}(\e-\w)\right) \\
& \qquad + \left(N^{\rm P}_\w - N^{\rm P}_{\w+eV}\right)
\left(\tilde{n}_{\rm r}(\e) - \tilde{n}_{\rm l}(\e-\w)\right) \Big]\bigg\}
\end{split}\ee
Thanks to the identity
\be\label{identity}\begin{split}
&\frac{\delta_1}{2\pi} \Re \left[(1+F^\rho){\cal L}^\rho_\w - {\cal L}^g_\w \right] \\
&\qquad = \frac{1}{\gch}
\frac{\w^2\tau^2F^\rho(2+F^\rho)}{(\w^2\tau^2+1)(\w^2\tau^2+(1+F^\rho)^2)},
\end{split}\ee
it becomes evident that
the correction to the distribution function $\delta n(\e)$
due to the electron-electron interaction has a small factor
$1/\gch$. All other factors in the expression for $\delta n(\e)$
are of the order of unity; in particular, the combinations of the
divergent Plank's functions $N^{\rm P}(\omega)$ and the Fermi
functions $n_{\rm F}(\e)$ are such that the integrand for
$\delta n(\e)$ does not have divergences at either
frequency $\omega=0, \ \pm eV$.

To arrive at the final expression for the correction to the noise power, we substitute
\reqs{itsol} and \rref{identity} into \req{noise} and obtain
\begin{subequations}
\label{main}
\begin{eqnarray}
\delta S &=&  2e{\cal F}|I| \frac{\chi(eV,T,F^\rho)}{\gch};\\
\label{maina}
\chi &=& \int\!
\frac{(F^\rho)^2 K_1(eV,\w)-2F^\rho K_2(eV,\w)}
{(\w^2\tau^2+1)(\w^2\tau^2+(1+F^\rho)^2)} \frac{d\w}{eV}\, , \qquad
\label{mainb}
\end{eqnarray}
\end{subequations}
where
\be\label{Ks}
\begin{split}
&\frac{K_1 (U,\w)}{\tau^2}
 =Y(U,\w)\!\left[\frac{Y(\w,0)}{2}+{\cal F}\big(Y(\w,U)-Y(\w,0)\big)\right],\\
&\frac{K_2(U,\w)}{\w\tau^2}
 =
\left[ \sum_\pm N^{\rm P}_{\w\pm U} -2N^{\rm P}_\w \right]
\left[Y(\w,U)+Y(\w,0)\right]
 \\
&\ \ \ \  + \left[\sum_\pm (\pm 1) N^{\rm P}_{\w\pm U} \right]
 \left[\sum_\pm (\pm 1) (\w\pm U)\coth\frac{\w\pm U}{2T} \right]  .
\end{split}
\ee
Equations \rref{main} and \rref{Ks} are the main result of the paper. They are valid
when random matrix theory is applicable and up to a large bias of order $V\sim \gch^2 E_C/e$;
above this bias, the assumption justifying the iterative solution in the form of \req{n0}
is not true anymore, although the kinetic equation approach is still viable.

The correction $\delta n(\e)$ to the electron distribution function in the leads
vanishes in equilibrium, when  $V=0$. In this case the
current noise contains only the Johnson-Nyquist component, which, in
its turn, depends only on the linear conductivity of the system and
for a dot with reflectionless point contacts has no interaction
corrections~\cite{PRL82BA,PRB69GZ,PRL94BLF}. At finite bias $V\neq 0$, the interaction correction to
the noise appears. Below, we investigate the properties of $\delta
S$ for various relations between temperature $T$, voltage $V$, and
interaction strength $F^\rho$.

We consider first the zero temperature limit $T=0$; in this case
$K_2$ vanishes, and performing the frequency integration in \req{main} we find
\be
\label{lowT}
\begin{split}
\chi = \frac{F^\rho(1-2{\cal F})}{2+F^\rho}&
\bigg[\ln \sqrt{\frac{1+(\tau eV)^2}
{1+(\tau_{\mathrm{c}} eV)^2}}
\\&
+
\frac{\arctan eV\tau}{eV\tau}-
\frac{\arctan eV\tau_{\mathrm{c}}}{eV\tau_{\mathrm{c}}}
\bigg],
\end{split}\ee
where $\tau_{\mathrm{c}}= \tau/(1+F^\rho)$.
For large $F^\rho\gg 1$ and intermediate values of bias $V$,
$1/\tau \ll eV \ll E_{\rm C}\gch$,
\req{lowT} can be approximated by \req{eq1}, while at larger bias, $eV\gg E_{\rm C}\gch$,
it has the asymptote
$
\chi = \left(1-2{\cal F}\right)\ln(4E_{\rm C}/\mls)
$.
Finally, at low bias, $eV\tau\ll 1$, we have
\be
\chi = \frac{1}{6}(1-2{\cal F})\left(\frac{F^\rho}{1+F^\rho}\right)^2 (eV\tau)^2 ,
\ee
whose quadratic dependence on the voltage can be explained by a standard phase-space
argument. As the bias grows, this argument breaks down since the energy window
to explore ($\sim eV$) becomes too big compared to the energy
correlation in electron spectrum ($\sim 1/\tau$).

At finite temperature, the interaction correction to the noise power
acquires a temperature dependence. Particularly, in the regime of
validity of \req{eq1}, when $1 \ll eV\tau \ll F^\rho$,
the lowest order correction to the noise can be estimated as
\be\label{tcorr}
\frac{\chi(T)-\chi(T=0)}{\chi(T=0)}=-\frac{\pi T \tau}{\ln |eV\tau|}
\ee
at small temperature $T\ll 1/\tau$.

\begin{figure}
\includegraphics[width=0.43\textwidth]{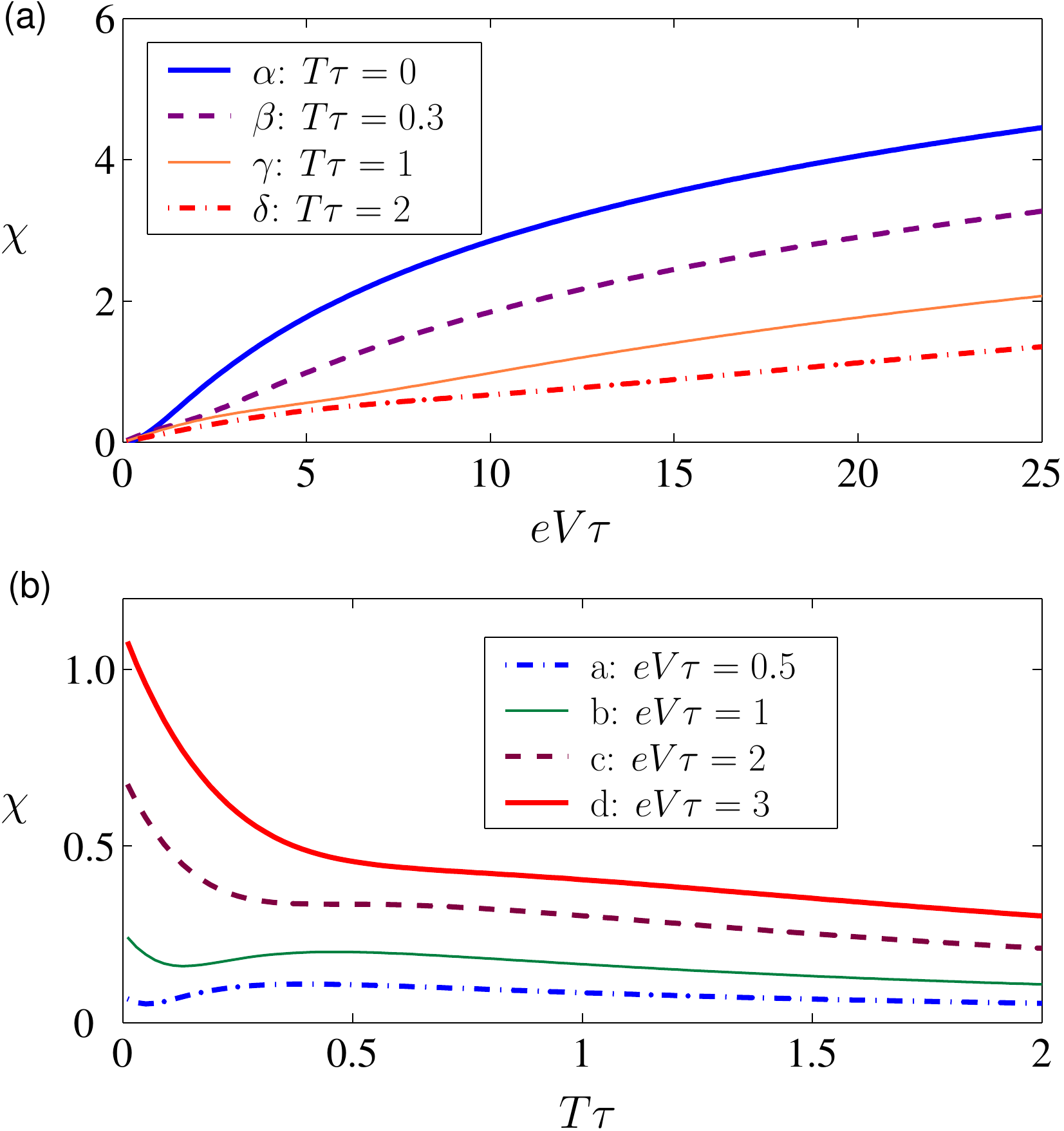}
\caption{(Color online). The correction to the noise power $\chi$ for $F^\rho=100$ (a) as a function
of bias for different temperatures and (b) as a function of temperature for different biases.
The labels refer to the cuts along the corresponding lines in Fig.~\ref{fig:1}(b).}
\label{fig:2}
\end{figure}

The dependence of $\chi(eV,T, F^\rho)$ on the charging energy
$E_{\rm C}=F^\rho\mls/4$ and bias $V$ is shown in Fig.~\ref{fig:1}(a)
for finite $T=1/\tau$. At small $F^\rho\lesssim 1$ and small bias
$eV\tau\lesssim 1$ the correction to the noise can be negative.
The negative values of $\chi$  at $F^\rho\lesssim 1$
originate from the $K_2$ term in \req{mainb}.  However, typically
$F^\rho\gg 1$, and the contribution of
the $K_1$ term dominates; the interaction correction to the
ensemble average noise power is then positive for any bias $V$, as
illustrated in Fig.~\ref{fig:2}(a), where dependence of $\chi$ on the
applied bias $V$ is shown for $F^\rho=100$.

We present in Fig.~\ref{fig:1}(b) the contour plot of $\chi(eV,T, F^\rho)$
as a function of bias $V$ and temperature $T$ at large $F^\rho=100$.
Although, in general, temperature suppresses the interaction correction
to the noise power, this suppression is nonmonotonic. This nonmonotonic
behavior is further illustrated in Fig.~\ref{fig:2}(b), where temperature
dependence of $\chi$ at fixed $V$ and $F^\rho$ is shown. The slope of the
initial temperature suppression can be calculated, at intermediate bias,
using \req{tcorr}.

Finally we note that in the limit of strong Coulomb interaction,
$F^\rho\gg 1$, the main contribution to the power of current noise
comes from term $K_1$ in \req{mainb}. This term originates from the
correction to the boson occupation number $N_\w$, which is given
by the second term in \req{Nalpha} containing $Y(eV,\w)$.
This nonequilibrium correction is due to the
coupling between the bosons and the electrons out of equilibrium.
If the boson occupation number $N_\w$ is not evaluated from the kinetic
equations, but chosen to describe the equilibrium state at the
temperature $T$ of the electrons in the leads, the leading order
contribution to the noise power vanishes.

In summary, we have considered the noise power of quantum dots with
reflectionless contacts, averaged over the unitary ensemble. For such
dots, the Coulomb interaction does not affect the average conductance,
but only its mesoscopic fluctuations. Here we have shown that the average
noise acquires an interaction correction in the nonequilibrium state.
This correction is present at zero temperature and is suppressed as
temperature increases; interestingly, the
suppression can be a nonmonotonic function of temperature.

We acknowledge correspondence with Y. Nazarov and D. Bagrets.

\bibliography{nqd1}

\begin{thebibliography}{24}
\expandafter\ifx\csname natexlab\endcsname\relax\def\natexlab#1{#1}\fi
\expandafter\ifx\csname bibnamefont\endcsname\relax
  \def\bibnamefont#1{#1}\fi
\expandafter\ifx\csname bibfnamefont\endcsname\relax
  \def\bibfnamefont#1{#1}\fi
\expandafter\ifx\csname citenamefont\endcsname\relax
  \def\citenamefont#1{#1}\fi
\expandafter\ifx\csname url\endcsname\relax
  \def\url#1{\texttt{#1}}\fi
\expandafter\ifx\csname urlprefix\endcsname\relax\def\urlprefix{URL }\fi
\providecommand{\bibinfo}[2]{#2}
\providecommand{\eprint}[2][]{\url{#2}}

\bibitem[{\citenamefont{Aleiner et~al.}(2002)\citenamefont{Aleiner, Brouwer,
  and Glazman}}]{ABG}
\bibinfo{author}{\bibfnamefont{I.~L.} \bibnamefont{Aleiner}},
  \bibinfo{author}{\bibfnamefont{P.~W.} \bibnamefont{Brouwer}},
  \bibnamefont{and} \bibinfo{author}{\bibfnamefont{L.}~\bibnamefont{Glazman}},
  \bibinfo{journal}{Phys. Rep.} \textbf{\bibinfo{volume}{358}},
  \bibinfo{pages}{309} (\bibinfo{year}{2002}).

\bibitem[{\citenamefont{Brouwer and Aleiner}(1999)}]{PRL82BA}
\bibinfo{author}{\bibfnamefont{P.~W.} \bibnamefont{Brouwer}} \bibnamefont{and}
  \bibinfo{author}{\bibfnamefont{I.~L.} \bibnamefont{Aleiner}},
  \bibinfo{journal}{Phys. Rev. Lett.} \textbf{\bibinfo{volume}{82}},
  \bibinfo{pages}{390} (\bibinfo{year}{1999}).

\bibitem[{\citenamefont{Golubev and Zaikin}(2004)}]{PRB69GZ}
\bibinfo{author}{\bibfnamefont{D.~S.} \bibnamefont{Golubev}} \bibnamefont{and}
  \bibinfo{author}{\bibfnamefont{A.~D.} \bibnamefont{Zaikin}},
  \bibinfo{journal}{Phys. Rev. B} \textbf{\bibinfo{volume}{69}},
  \bibinfo{pages}{075318} (\bibinfo{year}{2004}).

\bibitem[{\citenamefont{Brouwer et~al.}(2005)\citenamefont{Brouwer, Lamacraft,
  and Flensberg}}]{PRL94BLF}
\bibinfo{author}{\bibfnamefont{P.~W.} \bibnamefont{Brouwer}},
  \bibinfo{author}{\bibfnamefont{A.}~\bibnamefont{Lamacraft}},
  \bibnamefont{and}
  \bibinfo{author}{\bibfnamefont{K.}~\bibnamefont{Flensberg}},
  \bibinfo{journal}{Phys. Rev. Lett.} \textbf{\bibinfo{volume}{94}},
  \bibinfo{pages}{136801} (\bibinfo{year}{2005}).

\bibitem[{\citenamefont{Aleiner and Glazman}(1998)}]{PRB57AG}
\bibinfo{author}{\bibfnamefont{I.~L.} \bibnamefont{Aleiner}} \bibnamefont{and}
  \bibinfo{author}{\bibfnamefont{L.~I.} \bibnamefont{Glazman}},
  \bibinfo{journal}{Phys. Rev. B} \textbf{\bibinfo{volume}{57}},
  \bibinfo{pages}{9608} (\bibinfo{year}{1998}).

\bibitem[{\citenamefont{Blanter and Buttiker}(2000)}]{ButtikerBlanter-PReps}
\bibinfo{author}{\bibfnamefont{Y.~M.} \bibnamefont{Blanter}} \bibnamefont{and}
  \bibinfo{author}{\bibfnamefont{M.}~\bibnamefont{Buttiker}},
  \bibinfo{journal}{Phys. Rep.} \textbf{\bibinfo{volume}{336}},
  \bibinfo{pages}{1} (\bibinfo{year}{2000}).

\bibitem[{\citenamefont{Iannaccone et~al.}(1998)\citenamefont{Iannaccone,
  Lombardi, Macucci, and Pellegrini}}]{PRL80iannaccone}
\bibinfo{author}{\bibfnamefont{G.}~\bibnamefont{Iannaccone}},
  \bibinfo{author}{\bibfnamefont{G.}~\bibnamefont{Lombardi}},
  \bibinfo{author}{\bibfnamefont{M.}~\bibnamefont{Macucci}}, \bibnamefont{and}
  \bibinfo{author}{\bibfnamefont{B.}~\bibnamefont{Pellegrini}},
  \bibinfo{journal}{Phys. Rev. Lett.} \textbf{\bibinfo{volume}{80}},
  \bibinfo{pages}{1054} (\bibinfo{year}{1998}).

\bibitem[{\citenamefont{Oberholzer et~al.}(2001)\citenamefont{Oberholzer,
  Sukhorukov, Strunk, Sch\"onenberger, Heinzel, and Holland}}]{PRL86oberholzer}
\bibinfo{author}{\bibfnamefont{S.}~\bibnamefont{Oberholzer}},
  \bibinfo{author}{\bibfnamefont{E.~V.} \bibnamefont{Sukhorukov}},
  \bibinfo{author}{\bibfnamefont{C.}~\bibnamefont{Strunk}},
  \bibinfo{author}{\bibfnamefont{C.}~\bibnamefont{Sch\"onenberger}},
  \bibinfo{author}{\bibfnamefont{T.}~\bibnamefont{Heinzel}}, \bibnamefont{and}
  \bibinfo{author}{\bibfnamefont{M.}~\bibnamefont{Holland}},
  \bibinfo{journal}{Phys. Rev. Lett.} \textbf{\bibinfo{volume}{86}},
  \bibinfo{pages}{2114} (\bibinfo{year}{2001}).

\bibitem[{\citenamefont{Chen and Webb}(2006{\natexlab{a}})}]{PRB73CW}
\bibinfo{author}{\bibfnamefont{Y.}~\bibnamefont{Chen}} \bibnamefont{and}
  \bibinfo{author}{\bibfnamefont{R.~A.} \bibnamefont{Webb}},
  \bibinfo{journal}{Phys. Rev. B} \textbf{\bibinfo{volume}{73}},
  \bibinfo{pages}{035424} (\bibinfo{year}{2006}{\natexlab{a}}).

\bibitem[{\citenamefont{Chen and Webb}(2006{\natexlab{b}})}]{PRL97CW}
\bibinfo{author}{\bibfnamefont{Y.}~\bibnamefont{Chen}} \bibnamefont{and}
  \bibinfo{author}{\bibfnamefont{R.~A.} \bibnamefont{Webb}},
  \bibinfo{journal}{Phys. Rev. Lett.} \textbf{\bibinfo{volume}{97}},
  \bibinfo{pages}{066604} (\bibinfo{year}{2006}{\natexlab{b}}).

\bibitem[{\citenamefont{Barthold et~al.}(2006)\citenamefont{Barthold, Hohls,
  Maire, Pierz, and Haug}}]{PRL96barthold}
\bibinfo{author}{\bibfnamefont{P.}~\bibnamefont{Barthold}},
  \bibinfo{author}{\bibfnamefont{F.}~\bibnamefont{Hohls}},
  \bibinfo{author}{\bibfnamefont{N.}~\bibnamefont{Maire}},
  \bibinfo{author}{\bibfnamefont{K.}~\bibnamefont{Pierz}}, \bibnamefont{and}
  \bibinfo{author}{\bibfnamefont{R.~J.} \bibnamefont{Haug}},
  \bibinfo{journal}{Phys. Rev. Lett.} \textbf{\bibinfo{volume}{96}},
  \bibinfo{pages}{246804} (\bibinfo{year}{2006}).

\bibitem[{\citenamefont{Onac et~al.}(2006)\citenamefont{Onac, Balestro,
  Trauzettel, Lodewijk, and Kouwenhoven}}]{PRL96kouwenhoven}
\bibinfo{author}{\bibfnamefont{E.}~\bibnamefont{Onac}},
  \bibinfo{author}{\bibfnamefont{F.}~\bibnamefont{Balestro}},
  \bibinfo{author}{\bibfnamefont{B.}~\bibnamefont{Trauzettel}},
  \bibinfo{author}{\bibfnamefont{C.~F.~J.} \bibnamefont{Lodewijk}},
  \bibnamefont{and} \bibinfo{author}{\bibfnamefont{L.~P.}
  \bibnamefont{Kouwenhoven}}, \bibinfo{journal}{Phys. Rev. Lett.}
  \textbf{\bibinfo{volume}{96}}, \bibinfo{pages}{026803}
  (\bibinfo{year}{2006}).

\bibitem[{\citenamefont{Zhang et~al.}(2007)\citenamefont{Zhang, DiCarlo,
  McClure, Yamamoto, Tarucha, Marcus, Hanson, and Gossard}}]{0703419zhang}
\bibinfo{author}{\bibfnamefont{Y.}~\bibnamefont{Zhang}},
  \bibinfo{author}{\bibfnamefont{L.}~\bibnamefont{DiCarlo}},
  \bibinfo{author}{\bibfnamefont{D.~T.} \bibnamefont{McClure}},
  \bibinfo{author}{\bibfnamefont{M.}~\bibnamefont{Yamamoto}},
  \bibinfo{author}{\bibfnamefont{S.}~\bibnamefont{Tarucha}},
  \bibinfo{author}{\bibfnamefont{C.~M.} \bibnamefont{Marcus}},
  \bibinfo{author}{\bibfnamefont{M.~P.} \bibnamefont{Hanson}},
  \bibnamefont{and} \bibinfo{author}{\bibfnamefont{A.~C.}
  \bibnamefont{Gossard}} \bibinfo{journal}{Phys. Rev. Lett.}
  \textbf{\bibinfo{volume}{99}}, \bibinfo{pages}{036603}
  (\bibinfo{year}{2007}).

\bibitem[{\citenamefont{Beenakker and B\"uttiker}(1992)}]{PRB46BB}
\bibinfo{author}{\bibfnamefont{C.~W.~J.} \bibnamefont{Beenakker}}
  \bibnamefont{and}
  \bibinfo{author}{\bibfnamefont{M.}~\bibnamefont{B\"uttiker}},
  \bibinfo{journal}{Phys. Rev. B} \textbf{\bibinfo{volume}{46}},
  \bibinfo{pages}{1889} (\bibinfo{year}{1992}).

\bibitem[{\citenamefont{{de Jong} and Beenakker}(1997)}]{9611140dJB}
\bibinfo{author}{\bibfnamefont{M.~J.~M.} \bibnamefont{{de Jong}}}
  \bibnamefont{and} \bibinfo{author}{\bibfnamefont{C.~W.~J.}
  \bibnamefont{Beenakker}}, in \emph{\bibinfo{booktitle}{Mesoscopic Electron
  Transport}}, \bibinfo{series}{NATO Advanced Studies Institute, Series E},
  edited by L.L. Sohn, L.P. Kouwenhoven, G. Sch\"on 
  (\bibinfo{publisher}{Kluwer Academic Publishers, Dordrecht},
  \bibinfo{year}{1997}), Vol. \bibinfo{volume}{345}, p. \bibinfo{pages}{225}.

\bibitem[{\citenamefont{Bagrets and Nazarov}(2003)}]{PRB67BN}
\bibinfo{author}{\bibfnamefont{D.~A.} \bibnamefont{Bagrets}} \bibnamefont{and}
  \bibinfo{author}{\bibfnamefont{Y.~V.} \bibnamefont{Nazarov}},
  \bibinfo{journal}{Phys. Rev. B} \textbf{\bibinfo{volume}{67}},
  \bibinfo{pages}{085316} (\bibinfo{year}{2003}).

\bibitem[{\citenamefont{Bagrets and Nazarov}(2005)}]{PRL94BN}
\bibinfo{author}{\bibfnamefont{D.~A.} \bibnamefont{Bagrets}} \bibnamefont{and}
  \bibinfo{author}{\bibfnamefont{Y.~V.} \bibnamefont{Nazarov}},
  \bibinfo{journal}{Phys. Rev. Lett.} \textbf{\bibinfo{volume}{94}},
  \bibinfo{pages}{056801} (\bibinfo{year}{2005}).

\bibitem[{\citenamefont{Tobiska and Nazarov}(2005)}]{PRB72TN}
\bibinfo{author}{\bibfnamefont{J.}~\bibnamefont{Tobiska}} \bibnamefont{and}
  \bibinfo{author}{\bibfnamefont{Y.~V.} \bibnamefont{Nazarov}},
  \bibinfo{journal}{Phys. Rev. B} \textbf{\bibinfo{volume}{72}},
  \bibinfo{pages}{235328} (\bibinfo{year}{2005}).

\bibitem[{\citenamefont{Sch\"on and Zaikin}(1990)}]{PhysRep198SZ}
\bibinfo{author}{\bibfnamefont{G.}~\bibnamefont{Sch\"on}} \bibnamefont{and}
  \bibinfo{author}{\bibfnamefont{A.~D.} \bibnamefont{Zaikin}},
  \bibinfo{journal}{Phys. Reps.} \textbf{\bibinfo{volume}{198}},
  \bibinfo{pages}{237} (\bibinfo{year}{1990}).

\bibitem[{\citenamefont{Ahmadian et~al.}(2005)\citenamefont{Ahmadian, Catelani,
  and Aleiner}}]{aca}
\bibinfo{author}{\bibfnamefont{Y.}~\bibnamefont{Ahmadian}},
  \bibinfo{author}{\bibfnamefont{G.}~\bibnamefont{Catelani}}, \bibnamefont{and}
  \bibinfo{author}{\bibfnamefont{I.~L.} \bibnamefont{Aleiner}},
  \bibinfo{journal}{Phys. Rev. B} \textbf{\bibinfo{volume}{72}},
  \bibinfo{pages}{245315} (\bibinfo{year}{2005}).

\bibitem[{\citenamefont{Catelani and Vavilov}(2006)}]{cv-unp}
\bibinfo{author}{\bibfnamefont{G.}~\bibnamefont{Catelani}} \bibnamefont{and}
  \bibinfo{author}{\bibfnamefont{M.~G.} \bibnamefont{Vavilov}}
  (\bibinfo{note}{unpublished}).

\bibitem[{\citenamefont{Catelani and Aleiner}(2005)}]{ca}
\bibinfo{author}{\bibfnamefont{G.}~\bibnamefont{Catelani}} \bibnamefont{and}
  \bibinfo{author}{\bibfnamefont{I.~L.} \bibnamefont{Aleiner}},
  \bibinfo{journal}{Sov. Phys. JETP} \textbf{\bibinfo{volume}{100}},
  \bibinfo{pages}{331} (\bibinfo{year}{2005}) 
  [\bibinfo{note}{Zh. Eksp. Teor. Fiz. \textbf{127}, 372 (2005)}].

\bibitem[{\citenamefont{Schoelkopf et~al.}(1998)\citenamefont{Schoelkopf,
  Kozhevnikov, Prober, and Rooks}}]{PRL80SchProber}
\bibinfo{author}{\bibfnamefont{R.~J.} \bibnamefont{Schoelkopf}},
  \bibinfo{author}{\bibfnamefont{A.~A.} \bibnamefont{Kozhevnikov}},
  \bibinfo{author}{\bibfnamefont{D.~E.} \bibnamefont{Prober}},
  \bibnamefont{and} \bibinfo{author}{\bibfnamefont{M.~J.} \bibnamefont{Rooks}},
  \bibinfo{journal}{Phys. Rev. Lett.} \textbf{\bibinfo{volume}{80}},
  \bibinfo{pages}{2437} (\bibinfo{year}{1998}).

\bibitem[{\citenamefont{Levitov et~al.}(1996)\citenamefont{Levitov, Lee, and
  Lesovik}}]{JMatPhys37LLL}
\bibinfo{author}{\bibfnamefont{L.~S.} \bibnamefont{Levitov}},
  \bibinfo{author}{\bibfnamefont{H.}~\bibnamefont{Lee}}, \bibnamefont{and}
  \bibinfo{author}{\bibfnamefont{G.~B.} \bibnamefont{Lesovik}},
  \bibinfo{journal}{J. Math. Phys.} \textbf{\bibinfo{volume}{37}},
  \bibinfo{pages}{4845} (\bibinfo{year}{1996}).

\end{thebibliography}

\end{document}